\documentclass[9pt,twocolumn,twoside]{osajnl}

\usepackage{braket}
\let\smallket\ket
\let\smallbra\bra
\usepackage{physics}
\usepackage{bm}

\DeclareMathOperator{\pcmse}{PCMSE}

\DeclareMathOperator{\frc}{FRC}
\DeclareMathOperator{\pcfrc}{PCFRC}
\DeclareMathOperator{\re}{Re}

\newcommand{\mat}[1]{\bm{#1}}

\journal{ol} 

\setboolean{shortarticle}{true}

\title{Error metrics for partially coherent wavefields}

\author[1,*]{Abraham Levitan}
\author[1]{Riccardo Comin}

\affil[1]{Massachusetts Institute of Technology, 77 Mass Avenue, Cambridge, MA, USA}

\affil[*]{Corresponding author: alevitan@mit.edu}

\begin{abstract}
Lensless imaging methods that account for partial coherence have become very common in the past decade. However, there are no metrics in use for comparing partially coherent light fields, despite the widespread use of such metrics to compare fully coherent objects and wavefields. Here, we show how reformulating the mean squared error and Fourier ring correlation in terms of quantum state fidelity naturally generalizes them to partially coherent wavefields. These results fill an important gap in the lensless imaging literature and will enable quantitative assessments of the reliability and resolution of reconstructed partially coherent wavefields.
\end{abstract}

\setboolean{displaycopyright}{true}

\begin{document}

\maketitle

When demonstrating a new imaging method or reconstruction algorithm, it is important to quantitatively study its accuracy and reliability. In the ptychography literature, this is usually done by calculating a variation on the Mean Squared Error (MSE) \cite{fienup_invariant_1997} between reconstructed and ground truth objects, as a function of noise level or other parameters. Because the signal quality in an image is typically length-scale-dependent, it is also common to consider the resolution of an imaging method, i.e. the smallest length scale at which the retrieved image is sufficiently accurate. This is often accomplished with the Fourier Ring Correlation (FRC), which \cite{van_heel_similarity_1987} has the additional advantage that it can determine an empirical resolution from experimental data even when the ground truth is not known.

However, because of two recent trends these practices are no longer always sufficient. First, mixed state (or "multi-mode") ptychography methods have become popular since their introduction in 2013 \cite{thibault_reconstructing_2013}. These methods treat the illumination as an incoherent mixture of wavefields. In doing so, they account for the inevitable presence of partial coherence due to, for example, the intrinsic properties of the source \cite{cao_modal_2016, chen_mixed-state_2020, pound_ptychography_2020} or physical motion of the sample \cite{clark_continuous_2014, pelz_--fly_2014, deng_continuous_2015, huang_fly-scan_2015}.

Second, ptychography is now routinely used to characterize the probe itself, with applications ranging from understanding the statistical source properties of synchrotrons \cite{tsai_hard_2021} and free electron lasers \cite{sala_pulsetopulse_2019, daurer_ptychographic_2020, kharitonov_flexible_2021} to measuring the aberrations of x-ray optics \cite{huang_metrology_2020, takeo_soft_2020}. Therefore, it has become important to place the veracity of probes retrieved via mixed state ptychography on a firm footing. Because the the traditional metrics (MSE and FRC) only apply to fully coherent light, though, they cannot be used to study the consistency of multi-mode probe reconstructions.

Through a literature search we have identified two approaches which have been used to mitigate this problem in the absence of agreed-upon best practices. The first, most common method is to plot the orthogonalized modes of the light fields to be compared \cite{thibault_reconstructing_2013, cao_modal_2016, deng_continuous_2015, chen_mixed-state_2020, pound_ptychography_2020, tsai_hard_2021, sala_pulsetopulse_2019, daurer_ptychographic_2020}. This is comprehensive, but not quantitative. It is often augmented by a breakdown showing the relative power in each orthogonalized mode \cite{clark_continuous_2014, pelz_--fly_2014, huang_fly-scan_2015}. The mode breakdown can be used to generate quantitative comparisons, e.g. comparing the global degree of coherence between reconstruction attempts \cite{kharitonov_flexible_2021}, but these metrics are insensitive to variations in the spatial structure of the probe modes and thus cannot replace the traditional metrics.

The second, less common method is to compare each pair of orthogonalized probe modes in series using a metric such as the normalized MSE \cite{li_breaking_2016}. This takes into account the spatial structure of the probe, but it has the problem that the ordering of the modes can be unstable. There is also no natural way to reduce this list of comparisons to a single error metric, and it can't compare light fields with different numbers of modes.

A metric which could avoid these pitfalls should satisfy a few basic requirements:

\begin{enumerate}
\item It should be independent of the representation used, e.g. the ordering or number of modes.
\item It should be minimized only when comparing formally indistinguishable fields.
\item It should reduce to a metric already in widespread use when applied to coherent wavefields.
\end{enumerate}

The first condition is especially important because the multi-mode expansion is not the only way to treat partial coherence \cite{chang_partially_2018}, and any general solution to this issue must recognize that. This, together with the second condition, implies that such a metric should have a definition in terms of the density matrix $\mat{\rho} = \rho(\vec{r},\vec{r}')$ \cite{thibault_reconstructing_2013}, also known as the mutual coherence function \cite{wolf_new_1982, wolf_new_1986}. This is because $\mat{\rho}$ is the most general description of a monochromatic partially coherent wavefield, which all other representations can be rephrased in terms of.

However, to find a metric which meets condition 3 it will be helpful to link the density matrix representation to one defined explicitly in terms of coherent wavefields. We do this by exploring a common model of partial coherence, as the consequence of averaging over a time-varying coherent wavefield. Specifically, integrating the diffracted intensity from a time-varying wavefield $\ket{\psi(t)}$ over a period of time $T$ is equivalent to simulating that diffraction using the density matrix \cite{thibault_reconstructing_2013}

\begin{equation}
\mat{\rho} = \int_0^T dt \mat{\rho}(t) = \int_0^T dt \ket{\psi(t)}\bra{\psi(t)}.\label{eq:rhotopsi}
\end{equation}

Crucially, the time-dependent representation still describes a coherent wavefield, but with an extra dimension (time). This suggests that we might be able to generalize metrics from coherent to partially coherent wavefields by applying them in the time-dependent representation.

Initially this seems like a fool's errand because each density matrix $\mat{\rho}$ corresponds to infinitely many wavefields $\ket{\psi(t)}$. However, there will still be a unique minimum, corresponding to the best case which is consistent with the known information. In the case of the MSE calculated between two wavefields $\ket{\psi_1(t)}$ and $\ket{\psi_2(t)}$, we can set up the following minimization problem:

\begin{align}
\min_{\psi_1,\psi_2} \quad & \braket{\psi_1} + \braket{\psi_2} - 2 \re\left[\braket{\psi_1}{\psi_2}\right] \label{eq:msemin}\\
\text{s.t.} \quad & \mat{\rho_i} = \int dt \ket{\psi_i(t)}\bra{\psi_i(t)}, \quad i\in\{1,2\}, \nonumber
\end{align}

where the time-dependence is suppressed in the inner products to indicate that they integrate over time as well as the spatial/pixel dimensions. The first two terms are the traces ($\Tr$) of $\mat{\rho_1}$ and $\mat{\rho_2}$ respectively, and do not depend on the choice of $\ket{\psi_i(t)}$. Further, because $\ket{\psi_1(t)}$ and $\ket{\psi_2(t)}$ have a global phase degree of freedom, their overlap can be chosen real and non-negative. Therefore, \eqref{eq:msemin} simplifies to

\begin{equation}
\Tr(\mat{\rho_1}) + \Tr(\mat{\rho_2}) - 2 \max_{\psi_1,\psi_2} \left|\braket{\psi_1}{\psi_2} \right|
\end{equation}

With the maximization problem operating under the same constraints as \eqref{eq:msemin}. Happily, this problem has a well known solution. In the language of quantum states, $\ket{\psi_1(t)}$ and $\ket{\psi_2(t)}$ are purifications of the density matrices $\mat{\rho_1}$ and $\mat{\rho_2}$. A classic result in quantum information is that the maximum overlap between the purifications of a pair of density matrices is equal to their square-root fidelity ($\mathbb{F}$) \cite{uhlmann_transition_1976, jozsa_fidelity_1994}:

\begin{equation}
\mathbb{F}(\mat{\rho}_1,\mat{\rho}_2) = \Tr\left(\sqrt{\sqrt{\mat{\rho}_1} \mat{\rho}_2 \sqrt{\mat{\rho}_1}}\right) = \max_{\psi_1,\psi_2} \left|\braket{\psi_1}{\psi_2} \right|
\end{equation}

Consequently, the solution to \eqref{eq:msemin} (which we define as the partially coherent MSE, $\pcmse$) can be written directly in terms of the density matrices:

\begin{equation}
\pcmse(\mat{\rho}_1,\mat{\rho}_2) = \Tr(\mat{\rho}_1) + \Tr(\mat{\rho}_2) - 2\mathbb{F}(\mat{\rho}_1,\mat{\rho}_2).\label{eq:pcmse}
\end{equation}

As required, this reduces to the MSE maximized over the phase degree of freedom \cite{fienup_invariant_1997} when $\mat{\rho}_1$ and $\mat{\rho}_2$ represent pure states, and it depends only on the information in the density matrix representation. But, does it satisfy condition 2? It is straightforward to show that 

\begin{equation}
0 \leq \mathbb{F}(\mat{\rho}_1,\mat{\rho}_2) \leq \sqrt{\Tr(\mat{\rho}_1)\Tr(\mat{\rho}_2)} \label{eq:fidelitylim}
\end{equation}

with the upper equality achieved only when $\frac{\mat{\rho}_1}{\Tr(\mat{\rho}_1)} = \frac{\mat{\rho}_2}{\Tr(\mat{\rho}_2)}$ and the lower equality achieved only when $\mat{\rho}_1\mat{\rho}_2=0$ (see supplement 1.1). Consequently,

\begin{equation}
0 \leq \pcmse(\mat{\rho}_1,\mat{\rho}_2) \leq \Tr(\mat{\rho}_1) + \Tr(\mat{\rho}_2)
\end{equation}

with the lower equality achieved only when $\mat{\rho}_1 = \mat{\rho}_2$. Not only does this metric clearly satisfy condition 2, its bounds mirror those of the standard MSE. Consequently, we can generalize the normalized MSE by normalizing to $\Tr(\mat{\rho})$. It also is worth noting that the minimization of the normalized $\pcmse$ over a global amplitude degree of freedom has an especially simple definition, which parallels that derived in \cite{fienup_invariant_1997}:

\begin{equation}
\min_a \left[\frac{\pcmse(\mat{\rho}_1,a\mat{\rho}_2)}{\Tr(\mat{\rho}_1)}\right] = 1 - \frac{\mathbb{F}(\mat{\rho}_1,\mat{\rho}_2)^2}{\Tr(\mat{\rho}_1)\Tr(\mat{\rho}_2)}.
\end{equation}

This form, which is just $1$ minus the fidelity of the normalized density matrices, is very useful in ptychography where the intensity ratio of two probe reconstructions cannot always be determined.

The $\pcmse$ therefore seems like an ideal error metric, but based on the discussion so far it remains completely impractical to calculate. This is because the density matrices are usually so large that it is not even possible to hold them in memory, much less calculate their square roots. To make it practical, we need a way to calculate the $\pcmse$ directly from the multi-mode representation. In this representation, an $N\times M$ matrix $\mat{\psi}$ is stored, such that

\begin{equation}
\mat{\rho} = \mat{\psi}\mat{\psi}^\dagger. \label{eq:multimode}
\end{equation}

$N$ is the number of pixels in the image, and $M$ is the number of modes, i.e. the assumed maximum rank of $\mat{\rho}$. There are, unsurprisingly, strong connections between this expression and the time-dependent breakdown in \eqref{eq:rhotopsi}. To start, we note that $\Tr(\mat{\rho}) = \Tr(\mat{\psi}^\dagger\mat{\psi})$, the sum of the integrated intensities of each mode. Perhaps more surprisingly, the square root fidelity is also cheap to calculate,

\begin{equation}
\mathbb{F}(\mat{\rho}_1,\mat{\rho}_2) = ||\mat{\psi}^\dagger_1\mat{\psi}_2||_*
\end{equation}

where $||||_*$ is the nuclear norm, i.e. the sum of the singular values (see supplement 1.2). This form only requires a singular value decomposition of an $M_1\times M_2$ matrix. As a result, calculating the $\pcmse$ between two mixed state reconstructions can be done in the same amount of time it would take to calculate the MSE between $M_1 M_2$ pairs of coherent wavefields. Finally, because this form does not rely on any properties of $\mat{\psi}$ other than \eqref{eq:multimode} it still satisfies condition 1 - the two expansions need not even have the same number of modes! 

The derivations above constitute a complete framework for reducing the difference between partially coherent wavefields to a single number. However, there is also a need for metrics which can empirically assess the resolution of experimental results, i.e. an extension of the FRC. In practice, FRCs are rarely used to characterize retrieved probes, but it is our belief that as ptychography becomes a standard diagnostic of illumination sources the need for frequency-dependent analysis of the reliability of retrieved probe functions will grow.

The coherent FRC is calculated by splitting each of two images into a set of concentric rings in Fourier space, and calculating the correlation coefficient between each pair of rings. This curve, as a function of spatial frequency, is used to determine the length scale below which the images can no longer be considered reliable \cite{van_heel_similarity_1987, VanHeel2005}. We formalize this by defining a projection operator for each ring spanned by frequencies $k_1 < k_2$ in Fourier space:

\begin{equation} \mat{P}_{k_1,k_2} = \int d^2k \smallket{\vec{k}}\smallbra{\vec{k}} \theta(|\vec{k}| - k_1) \theta(k_2 - |\vec{k}|).
\end{equation}

Here $\smallket{\vec{k}}$ is a complex exponential at frequency $\vec{k}$ and $\theta$ is the Heaviside step function. We can then define the FRC for coherent fields as

\begin{equation} \frc(\ket{\psi_1},\ket{\psi_2}; k_1,k_2) = \frac{\bra{\psi_1}\mat{P}\ket{\psi_2}}{\sqrt{\bra{\psi_1}\mat{P}\ket{\psi_1}\bra{\psi_2}\mat{P}\ket{\psi_2}}},\label{eq:frc}
\end{equation}

where the subscript $k_1,k_2$ is dropped from the projection operators for compactness. We should quickly note that the FRC, as originally envisioned, is a real function of real-valued fields which can be positive or negative. However, for complex-valued fields, it is complex. Universally, what is reported in the ptychography literature is the magnitude of the FRC, i.e. the maximum over a phase degree of freedom.

Generalizing this definition to density matrices in the vein of our earlier approach means maximizing \eqref{eq:frc} over the space of all consistent purifications. This results in a simple expression for the Partially Coherent FRC ($\pcfrc$) as the normalized Fidelity of the projected density matrices:

\begin{equation}
\pcfrc(\mat{\rho}_1,\mat{\rho}_2; k_1,k_2) = \frac{\mathbb{F}(\mat{P}\mat{\rho}_1\mat{P},\mat{P}\mat{\rho}_2\mat{P})}{\sqrt{\Tr(\mat{P}\mat{\rho}_1\mat{P})\Tr(\mat{P}\mat{\rho}_2\mat{P})}}.
\end{equation}

This is not surprising because the FRC is essentially a normalized overlap, and the Fidelity is just the generalization of the overlap to mixed states. It is evident from \eqref{eq:fidelitylim} that this expression is constrained between $0$ and $1$, just like the standard FRC. In addition, because the projection operators are easy to implement in the multi-mode framework, the $\pcfrc$ remains cheap to calculate, equivalent to the cost of calculating $M_1 M_2$ coherent FRCs.

\begin{figure}[htbp]
\centering
\fbox{\includegraphics[width=\linewidth]{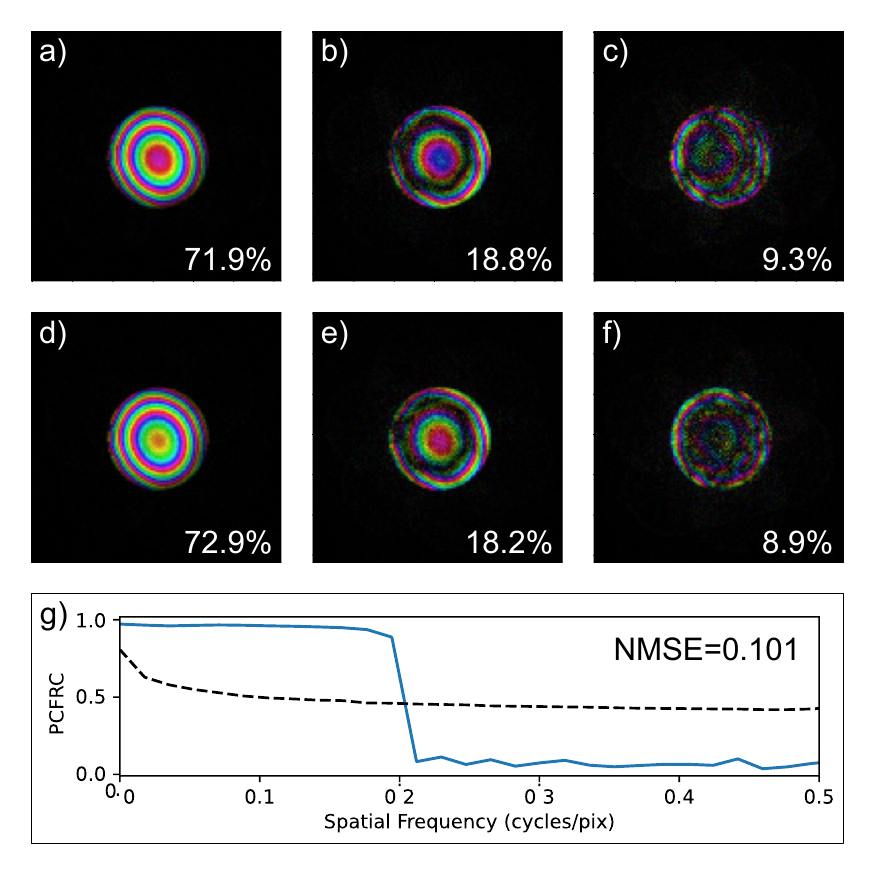}}
\caption{Example use of the $\pcfrc$ and $\pcmse$ to estimate the reliability of a probe reconstruction. (a-c) and (d-f) are two three-mode probe reconstructions from separate halves of a reference dataset from Chen et al. \cite{chen_mixed-state_2020}. The probes are shown in Fourier space to emphasize the connection with the $\pcfrc$. (g) contains the $\pcfrc$ between the two reconstructed partially coherent probes, compared to a half-bit threshold \cite{VanHeel2005}.}
\label{fig:frcimage}
\end{figure}

Finally, to demonstrate how these metrics perform in a typical case, in Figure \ref{fig:frcimage} we show a pair of reconstructed probes from two halves of the electron ptychography dataset made publicly available by Chen et al. \cite{chen_mixed-state_2020}. The reconstructions remain nonzero up to a maximum frequency defined by the condenser's aperture. This is reflected in the $\pcfrc$, which falls sharply from a high value near $1$ within the aperture to a low value near zero beyond it. The amplitude-minimized normalized $\pcmse$ is $0.101$, indicating that although the reconstructions are visually similar, a wavefield containing at least $10.1\%$ of the power in the first probe would be needed to map it onto the second probe. Further analysis of the $\pcmse$ and $\pcfrc$ using this dataset, involving comparisons between reconstructions with different numbers of modes, is reported in supplement 1.3.

In sum, we have generalized the MSE and FRC from coherent images to the space of partially coherent wavefields by finding an analogy with the quantum state fidelity. The results reduce to simple expressions which are computationally cheap when the density matrices are stored in the low rank multi-mode approximation. These metrics address a major need in the coherent imaging community for quantitative analysis of reconstructed partially coherent wavefields, and we hope they will find widespread use as the computational study of partial coherence continues to expand.

\begin{backmatter}
\bmsection{Funding} This material is based upon work supported by the Department of Energy, Office of Science, Office of Basic Energy Sciences, under Award Number DE-SC0021939.

\bmsection{Acknowledgments} The authors are deeply thankful for Kahraman Keskinbora's comments on an early version of this manuscript.

\bmsection{Disclosures} The authors declare no conflicts of interest.

\bmsection{Data availability} Only datasets previously made public were analyzed in this paper. A python implementation of the $\pcmse$ and $\pcfrc$ will be made available upon publication at \href{https://doi.org/10.7910/DVN/CUTW1U}{doi:10.7910/DVN/CUTW1U}.

\bmsection{Supplemental document}
See Supplement 1 for full derivations. 

\end{backmatter}

\bibliography{bibliography}

\bibliographyfullrefs{bibliography}


\end{document}


\maketitle

\section{Bounds on Fidelity and PCMSE}

For any two matrices with trace 1, the bound

\[ 0 \leq \mathbb{F}(\mat{\rho}_1,\mat{\rho}_2) \leq 1 \]

is well established \cite{jozsa_fidelity_1994}, with equality only when $\mat{\rho}_1 = \mat{\rho}_2$. For any matrix $\mat{\rho}$ with arbitrary trace, the matrix $\frac{\mat{\rho}}{\Tr(\mat{\rho})}$ has trace 1, so

\[ 0 \leq \mathbb{F}\left(\frac{\mat{\rho}_1}{\Tr(\mat{\rho}_1)},\frac{\mat{\rho}_2}{\Tr(\mat{\rho}_2)}\right) \leq 1. \]

From the definition of the square-root fidelity $\mathbb{F}$, it is clear that

\[ \mathbb{F}(a \mat{\rho}_1,b\mat{\rho}_2) = \sqrt{ab} \ \mathbb{F}( \mat{\rho}_1,\mat{\rho}_2)  \]

for scalar $a,b$. Therefore, we find 

\[ 0 \leq \mathbb{F}(\mat{\rho}_1,\mat{\rho}_2) \leq 
\sqrt{\Tr(\mat{\rho}_1)\Tr(\mat{\rho}_2)}. \]

Plugging this into the expression for the PCMSE, we find

\[ \Tr(\mat{\rho}_1) + \Tr(\mat{\rho}_2) \geq \pcmse (\mat{\rho}_1,\mat{\rho}_2) \geq \Tr(\mat{\rho}_1) + \Tr(\mat{\rho}_2) - 2 * \sqrt{\Tr(\mat{\rho}_1)\Tr(\mat{\rho}_2)}. \]

Finally, as a consequence of the AM-GM inequality

\[ \Tr(\mat{\rho}_1) + \Tr(\mat{\rho}_2) \geq \pcmse (\mat{\rho}_1,\mat{\rho}_2) \geq 0, \]

with the lower bound holding only when $\Tr(\mat{\rho}_1) = \Tr(\mat{\rho}_2)$ and $\frac{\mat{\rho_1}}{\Tr(\mat{\rho_1})} = \frac{\mat{\rho_2}}{\Tr(\mat{\rho_2})}$, i.e. $\mat{\rho_1} = \mat{\rho_2}$ 

\section{Multi-mode expression for Fidelity}

We expand the definition of fidelity using the multi-mode representation of $\mat{\rho}_2$:

\[ \mathbb{F}(\mat{\rho}_1,\mat{\rho}_2) = \Tr\left(\sqrt{\sqrt{\mat{\rho}_1} \mat{\psi}_2 \mat{\psi}_2^\dagger \sqrt{\mat{\rho}_1}}\right) \]

Noting that $\sqrt{\mat{\rho}_1}$ is Hermetian, we can see that this expression is equivalent to

\[ \mathbb{F}(\mat{\rho}_1,\mat{\rho}_2) = ||\sqrt{\mat{\rho}_1} \mat{\psi}_2||_* \]

by the definition of the nuclear norm. However, it is also the case that

\[ \sqrt{\mat{\rho}_1} \left(\sqrt{\mat{\rho}_1}\right)^\dagger = \mat{\rho}_1 = \mat{\psi}_1 \mat{\psi}_1^\dagger\]

Therefore, there must exist a semi-unitary matrix $\mat{U}$ satisfying $\mat{U}^\dagger \mat{U} = \mat{I}$ such that

\[ \left(\sqrt{\mat{\rho}_1}\right) = \left(\sqrt{\mat{\rho}_1}\right)^\dagger = U^\dagger \mat{\psi}_1^\dagger, \]

from \cite{horn_matrix_1985} Theorem (7.3.11). Therefore,

\[ \mathbb{F}(\mat{\rho}_1,\mat{\rho}_2) = ||U^\dagger \mat{\psi}_1^\dagger  \mat{\psi}_2||_* = ||\mat{\psi}^\dagger_1\mat{\psi}_2||_*, \]

due to the unitary invariance of the nuclear norm.

\section{Comparing reconstructions with varying numbers of modes}

\begin{figure}[htbp]

\centering
\fbox{\includegraphics[width=\linewidth]{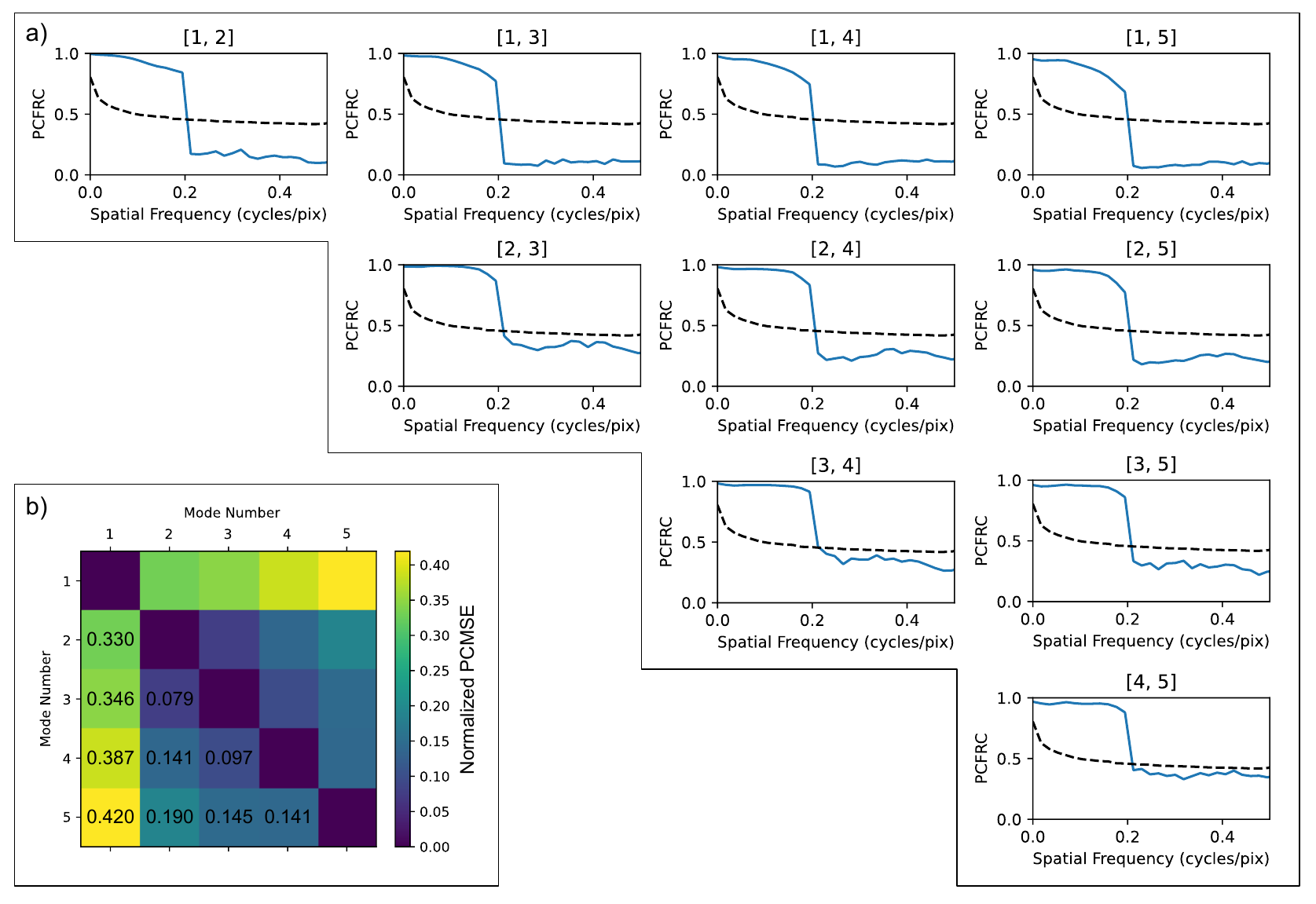}}
\caption{Comparisons between reconstructions with varying numbers of modes. (a) shows comparisons between each pair of reconstructions using the $\pcfrc$. (b) compares the same reconstructions using the amplitude-minimized normalized $\pcmse$.}
\label{fig:modecomp}
\end{figure}

Because the $\pcmse$ and $\pcfrc$ allow for comparisons between reconstructions of probe functions with different numbers of modes, we confirmed that comparisons perform as expected on typical experimental data. Therefore, we performed 5 reconstructions on the full dataset from reference \cite{chen_mixed-state_2020} which was used to generate Figure 1 of the main paper. The reconstructions were run with a varying number of modes from 1 to 5.

In Figure \ref{fig:modecomp}, we compare the partially coherent probes resulting from each reconstruction using the amplitude-minimized normalized $\pcmse$ and the $\pcfrc$. The results conform to our expectations. First, the introduction of even a single additional mode causes a dramatic change in the reconstructed probe as measured by $\pcmse$. This is expected because, in a two-mode reconstruction, the subdominant mode contains significant power ($23.5\%$). However, the mode-power breakdown hides the fact that the dominant mode in the two-mode reconstruction also has a different spatial structure from the single-mode result. This difference is captured by the normalized $\pcmse$ of $0.330$, which is higher than the power fraction in the subdominant mode due to the differing spatial structure of the two results.

Continuing on to higher modes, we find that the relative error introduced by each additional mode decreases until the fourth mode is added, at which point it begins to increase again. This is likely caused by higher order modes accumulating artifacts. This effect is again hidden by the orthogonalized mode power breakdown. For example, even though the fourth mode in a four-mode reconstruction only contains $5.1\%$ of the power, the normalized $\pcmse$ between the three- and four-mode reconstructions is nearly double at $0.97$. This again indicates that the structure of the lower modes has been modified to accommodate the additional mode, an effect which the $\pcmse$.

Finally, looking at the $\pcfrc$s reveals more information about the spatial structure of these changes. Notably, the $\pcfrc$s calculated between the one-mode and remaining results show a dip near the edge of the probe's Fourier space representation. This region corresponds to the the edge of the aperture in the electron microscope's condenser. The dip is therefore quantifying our expectation that the effects of incoherence are magnified near the edges of the condenser aperture, a well known phenomenon. As the number of modes increases, the edge of the $\pcfrc$ curve becomes more defined. This indicates that although minor changes are occurring as each mode is added, the essential structure of the probe has been sufficiently captured.

\bibliography{bibliography}
